\documentclass[usletter,11pt]{article}
\usepackage{fullpage}
\usepackage{times}
\usepackage{amssymb}
\newtheorem{theorem}{Theorem}
\newtheorem{proposition}{Proposition}

\newtheorem{lemma}{Lemma}

\newtheorem{corollary}{Corollary}

\newtheorem{definition}{Definition}
\title{Computing multiway cut within the given excess over the largest minimum isolating cut}
\author{Igor Razgon\\ \small Department of Computer Science, University of Leicester\\ \small ir45@mcs.le.ac.uk}
\date{}
\begin{document}
\maketitle
\begin{abstract}
Let $(G,T)$ be an instance of the (vertex) multiway cut problem where $G$ is a graph
and $T$ is a set of terminals. For $t \in T$, a set of nonterminal vertices separating
$t$ from $T \setminus \{T\}$ is called an \emph{isolating cut} of $t$. The largest
among all the smallest isolating cuts is a natural lower bound for a multiway
cut of $(G,T)$. Denote this lower bound by $m$ and let $k$ be an integer.

In this paper we propose an $O(kn^{k+3})$ algorithm that computes a multiway cut
of $(G,T)$ of size at most $m+k$ or reports that there is no such multiway cut.
The core of the proposed algorithm is the following combinatorial result.
Let $G$ be a graph and  let $X,Y$ be two disjoint subsets of vertices of $G$.
Let $m$ be the smallest size of a vertex $X-Y$ separator. Then, for the given
integer $k$, the number of \emph{important} $X-Y$ separators \cite{MarxTCS} 
of size at most $m+k$ is at most $\sum_{i=0}^k{n \choose i}$. 
\end{abstract}
\section{Introduction}
{\bf 1.1. Results and motivation.} Let $(G,T)$ be a pair where $G$ is a graph and $T$ a subset of $V(G)$.
Let us call the vertices of $T$ the \emph{terminals}. A \emph{multiway cut}
of $(G,T)$ is a set $S$ of non-terminal vertices such that in $G \setminus S$
no two terminals belong to the same connected component. The multiway cut problem
\textsc{mwc} asks for the smallest multiway cut of $(G,T)$. For two terminals
this problem can be solved by network flow techniques but becomes NP-hard
for 3 terminals \cite{Dahlapprox}.

Let $t \in T$. An \emph{isolating cut} of $t$ \cite{Dahlapprox} is a set 
$S \subseteq V(G) \setminus T$ separating $t$ from the rest of terminals. 
Denote by $m(t)$ the size of the smallest isolating cut of $t$ and let
$m=max_{t \in T} m(t)$. It is not hard to see that $m$ is a \emph{polynomially
computable lower bound} on the size of the smallest multiway cut of $(G,T)$.

In this paper, we investigate computing a multiway cut with a bounded \emph{excess}
over $m$. In particular, the main result of this paper is an $O(kn^{k+3}+|T|n^3)$ time 
algorithm that checks whether $(G,T)$ has 
a solution of size at most $(m+k)$ for the given integer $k$.

The main motivation of the above result comes from Parameterized Complexity \cite{Grohebook}.
The \textsc{mwc} problem is well-known to be Fixed-Parameter Tractable (FPT)
parameterized by the solution size \cite{MarxTCS,ChenAlgorithmica}. 
Can we provide a better parameterization that addresses the case where the optimal solution 
is large? For some problems, positive answers have been obtained by introducing so called
\emph{parameterization above guaranteed value}, the idea first proposed in  \cite{RamanAbove}. 
We apply this template to the \textsc{mwc} problem. In our case,
the guaranteed value is $m$ (because the solution size is always $m$ or greater)
and we study the parameterization of the \textsc{mwc} problem by the excess over $m$.
The proposed result makes a progress in this study because it shows that the \textsc{mwc}
problem with respect to the considered parameter is in XP and this makes meaningful
the question as to \emph{whether the \textsc{mwc} problem is FPT parameterized by
the excess over the maximal size of a smallest isolating cut}. To the best of our knowledge
this is the first result result addressing the \textsc{mwc} problem parameterized
above a guaranteed value. 

The key ingredient in the proof of the above result is a combinatorial theorem bounding
the number of \emph{important} $X-Y$ separators \cite{MarxTCS} of excess at most $k$ over
the smallest one.
Let $X$ and $Y$ be two disjoint subsets of $V(G)$. 
Let $r$ be the size of a smallest $X-Y$ separator. It is known \cite{MarxTCS} that
there is exactly one important $X-Y$ separator of size $r$. But how many are there 
important separators of size at most $r+k$ for the given integer $k$? The best existing bound
is $4^{r+k}$ \cite{LokshtanovClustering,Multicut}. We prove that the number of
such important separators is at most $\sum_{i=0}^k{n \choose i}$, which is much better 
than $4^{r+k}$ if $r$ is large. 
To the best of our knowledge, this is the first upper bound on the number of important 
separators where the size of a separator is not in the exponent. This upper bound is obtained 
by observing that important separators have a number of nice structural properties that establish 
an injective function from the set of important separators of size at most $r+k$ to the family 
of subsets of vertices of size at most $k$.
\\
{\bf 1.2. Related work.}
The \textsc{mwc} problem is a natural generalization of the standard $s-t$ cut problem
having applications related to resource allocation such as Multiprocessor Scheduling \cite{StoneApplication}
and Medical Imaging \cite{Zabih1,MedImaging}. This problem has been shown NP-hard in \cite{Dahlapprox} even
for the case of three terminals. This gave rise
to the investigation of methods of coping with NP-hardness for the \textsc{mwc} problem.
In the direction of identifying polynomially solvable subclasses, the researchers mainly
concentrated on planarity and tree-like structures (e.g. \cite{MwayCutTrees,Dahlapprox,Hartvigsen}). 
Approximation algorithms for this problem have been also actively investigated resulting 
in a row of improvements and generalizations (see e.g. \cite{Dahlapprox,Rabaniapprox,Gargapprox}).

The notion of \emph{isolating cut} (for the edge \textsc{mwc} problem)
has been coined first in \cite{Dahlapprox} in connection to the design of an approximation 
algorithm. In \cite{Gargapprox}, the notion has been reformulated in terms of the vertex \textsc{mwc}
problem in the way used in the present paper. However, in \cite{Gargapprox} it is pointed out
that their algorithm is not based on this notion.

The parameterized version of the \textsc{mwc} problem was first considered in \cite{MarxTCS},
the solution size being the parameter. An algorithm with a significantly improved runtime
has been proposed in \cite{ChenAlgorithmica}. The key theorem behind this algorithm gave
rise to first FPT algorithms for the Directed Feedback Vertex Set \cite{DFVSalgo} and Min 2-CNF
deletions problems \cite{ROJCSS}, whose fixed-parameter tractabilities were long standing open
questions. We believe this is an indication that the \textsc{mwc} problem is a very convenient
framework for studying graph separation problems in the sense that it reveals some structural
properties relevant to many other problems but not easily seen there.

The notion of \emph{important separator} has been coined in \cite{MarxTCS}. 
It is explicitly used in \cite{LokshtanovClustering} and \cite{Multicut} for resolving
a number of challenging open problems. In fact, as pointed out in \cite{LokshtanovClustering},
\cite{ChenAlgorithmica,DFVSalgo,ROJCSS} also implicitly use important separators. This shows that
an important separator is an interesting an worth studying combinatorial concept.

Finally, the investigation of parameters above and below guaranteed values has been
initiated in \cite{RamanAbove}. Currently, it is an active research area. An overview
of it can be found in the introduction of \cite{GutinESA}. 
\\
{\bf 1.3. Structure of the paper.}
Section 2 introduces the necessary background notions and their basic properties. 
Section 3 introduces the notion of \emph{important witness}, a special
case of important separator, having some nice properties. Using these
properties, Section 4 shows that any non-smallest important separator is nothing else
but a \emph{compound witness}, a generalization of an important witness
uniquely associated with  a subset of vertices of size not greater than
its excess. From this the desired upper bound on the number of important
separators is derived and applied to the \textsc{mwc} problem.

\section{Preliminaries}
We employ a standard notation related to graphs. In particular, given a graph $G$, let $C \subseteq V(G)$.
Then $G[C]$ denotes the subgraph of $G$ induced by $C$ and $G \setminus C \equiv G[V(G) \setminus C]$.
For $v \in V(G)$, $G \setminus v \equiv G[V(G) \setminus \{v\}]$ and $N(v)$ is the set of neighbors of
$v$ in $G$. Also, $N(C) \equiv (\bigcup_{v \in C} N(v)) \setminus C$.

Let $X$ and $Y$ be two disjoint sets of vertices of the given graph $G$.
A set $K \subseteq V(G) \setminus (X \cup Y)$ is an $X-Y$ separator
if in $G \setminus K$ there is no path from $X$ to $Y$.
Let $A,B$ be two disjoint subsets of $V(G)$. We denote
by $NR(G,A,B)$ the set of vertices that are not reachable from $A$ in $G \setminus B$
Let $K_1$ and $K_2$ be two $X-Y$ separators. We say that $K_1 \geq K_2$ if
$NR(G,Y,K_1) \supseteq NR(G,Y,K_2)$.

\begin{proposition} \label{internal}
Let $K_1$ and $K_2$ be two minimal $X-Y$ separators. Then  $K_1 \leq K_2$
if and only if $K_1 \setminus K_2 \subseteq NR(G,Y,K_2)$.
\end{proposition}

{\bf Proof.}
Assume first that $K_1 \leq K_2$. Due to the minimality of $K_1$,
each $v \in K_1$ is adjacent to some vertex $w$ of $NR(G,Y,K_1)$.
Since $w \in NR(G,Y,K_2)$ by our assumption, $v \in NR(G,Y,K_2)$
whenever $v \in K_1 \setminus K_2$. For the opposite direction,
any vertex of $NR(G,Y,K_1)$ can be connected to $Y$ only through
$K_1$. Since in $G \setminus K_2$, all vertices of $K_1 \setminus K_2$
are disconnected from $Y$ such connection is impossible. $\blacksquare$

Let $K_1$ and $K_2$ be two minimal $X-Y$ separators.
Let $K_1^t=K_1 \cap NR(G,Y,K_2)$,
$K_1^b=(K_1 \setminus K_1^t) \setminus (K_1 \cap K_2)$. Accordingly, let $K_2^t=K_2 \cap NR(G,Y,K_1)$
and $K_2^b=(K_2 \setminus K_2^t) \setminus (K_1 \cap K_2)$ (the superscripts 't' and 'b' correspond to
the words 'top' and 'bottom'). We denote $K_1^t \cup K_2^t \cup (K_1 \cap K_2)$ and
$K_1^b \cup K_2^b \cup (K_1 \cap K_2)$ by, respectively, $Top_{G,X,Y}(K_1,K_2)$ and $Bottom_{G,X,Y}(K_1,K_2)$,
the subscripts may be omitted if they are clear from the context.

\begin{proposition} \label{kb}
Let the notation be as in the previous paragraph.
Then both $Top(K_1,K_2)$ and $Bottom(K_1,K_2)$ are $X-Y$ separators.
Moreover, $Bottom(K_1,K_2) \geq K_1$ and $Bottom(K_1,K_2) \geq K_2$.
\end{proposition}

{\bf Proof.}
Consider the set $N^*=NR(G,Y,K_1) \cup NR(G,Y,K_2)$. By definition of $K_1$
and $K_2$ this set includes $X$ and does not contain any vertex of $Y$. 
What is the set of neighbors of this set, i.e what is the set separating
$N^*$ from the rest of the graph? Clearly, it is a subset of $K_1 \cup K_2$
excluding those vertices that belong to $NR(G,Y,K_1) \cup NR(G,Y,K_2)$.
In other words, it is a subset of $Bottom(K_1,K_2)$, and no vertex of $Bottom(K_1,K_2)$
belongs to $N^*$. It follows that $Bottom(K_1,K_2)$ is $X-Y$ separator, separating from $Y$
a superset of $NR(G,Y,K_1)$ and of $NR(G,Y,K_2)$, i.e. 
$Bottom(K_1,K_2) \geq K_1$ and $Bottom(K_1,K_2) \geq K_2$ as required. $\blacksquare$

A minimal $X-Y$ separator $K$ is called \emph{important} if there is no $X-Y$ separator $K'$
such that $K < K'$ and $|K| \geq |K'|$. This notion was first introduced in \cite{MarxTCS}
in a slightly different form. In particular, let $R(G,X,K)$ be the set of vertices that belong
to the same component in $G \setminus K$ with at least one vertex of $X$. In the definition
of \cite{MarxTCS}, the condition $K<K'$ is replaced by $R(G,X,K) \subset R(G,X,K')$.
The following proposition shows that these conditions are equivalent thus implying 
the equivalence of definitions.

\begin{proposition} \label{equivalence}
Let $K$ and $K'$ be two distinct $X-Y$ separators of $G$.
Then $NR(G,Y,K) \subset NR(G,Y,K')$ if and only if $R(G,X,K) \subset R(G,X,K')$.
\end{proposition}
{\bf Proof.}
It is not hard to see that since $K \neq K'$, $NR(G,Y,K) \neq NR(G,Y,K')$
and $R(G,X,K) \neq R(G,X,K')$. Indeed, if $K$ is a minimal separator then
$K$ is the neighborhood of both $NR(G,Y,K)$ and $R(G,X,K)$, the same is,
of course true for $K'$. But the same set cannot have two different neighborhoods.
It follows that we can replace '$\subset$' by '$\subseteq$'
in the statement of the observation. Assume that  $NR(G,Y,K) \subseteq NR(G,Y,K')$
and let $v \in R(G,X,K)$. Then there is a $X-v$ path $p$ all vertices of which
belong to $R(G,X,K) \subseteq NR(G,Y,K) \subseteq NR(G,Y,K')$. It follows that
$v$ is reachable from $X$ in $G \setminus K'$, i.e. $v \in R(G,X,K')$. Conversely,
assume that $R(G,X,K) \subseteq R(G,X,K')$. Due to the minimality of $K$, each
$v \in K$ is adjacent to a component $C$ of $G \setminus K$ containing at least
one vertex of $X$. Since all the vertices of $C$ are preserved in $R(G,X,K')$,
$v \in R(G,X,K') \subseteq NR(G,Y,K')$ whenever $v \in K \setminus K'$. The 
desired statement now follows from Proposition \ref{internal}. $\blacksquare$

\begin{corollary} \label{onesmallest}
Let $r$ be the size of a smallest $X-Y$ separator of $G$.
Then there is exactly one important $X-Y$ separator $K$ of size $r$.
Moreover, $K^*>K$ for any other important separator $K^*$.
\end{corollary}

{\bf Proof.}
Having in mind Proposition \ref{equivalence}, the first statement
is Lemma 3.3. of \cite{MarxTCS} and the second statement (in fact,
both of them) are proven in the second and third paragraphs of the
proof of Lemma 2.6. of \cite{Multicut}. $\blacksquare$

For the result proposed in this paper, we will need to compute
the unique smallest important $X-Y$ separator. It is known to be
polynomially computable, see, for example Lemma 3.2. of \cite{MarxTCS}
for a more general polynomial computability statement. In the following
lemma, we show that computing the smallest important $X-Y$ separator
in fact takes the same time as computing an arbitrary smallest $X-Y$
separator.

\begin{lemma} \label{ComputeImport}
The smallest important $X-Y$ separator can be computed in
$O(n^3)$ by an algorithm that first computes in $O(n^3)$ a 
largest set of internally vertex disjoint $X-Y$ paths and then
spends additional $O(n^2)$ time to computing the smallest 
important $X-Y$ separator. 
\end{lemma}

{\bf Proof.}
Let $p_1, \dots, p_r$ be a largest set of internally
vertex disjoint $X-Y$ paths that can be computed in $O(n^3)$
using standard network flow techniques (the computation takes at
most $n+1$ iterations of Ford-Fulkerson algorithm each taking $O(n^2)$,
see, for example \cite{Cormen}). We are going to show
how to compute the smallest important $X-Y$ separator having
these paths computed. Assume that each $p_i$ is of length $r_i$
and enumerate its vertices $v_{i,1}, \dots, v_{i,r_i}$ in the order
they occur is $p_i$ being explored from $X$ to $Y$. We may assume
that for each $p_i$ $v_{i,1}$ is the only vertex of $X$ and $v_{i,r_i}$
is the only vertex of $Y$ otherwise we can just shorten these paths
to obtain the desired effect. We can also assume that $X$ and $Y$ 
are singletons $\{x\}$ and $\{y\}$, respectively: for the purpose of 
the considered problem $X$ and $Y$ can be safely contracted into single
vertices.

We use the concept of \emph{torso} introduced in \cite{TWRedSTACS}.
Recall that for $S \subseteq V(G)$, $torso(G,S)$ is the graph obtained
from $G[S]$ by introducing new edges between those vertices $v_1,v_2$
of $S$ that are connected by path all intermediate vertices of which lie 
outside $S$. Denote $V(p_1) \cup \dots V(p_r)$ by $V^*$ and consider the
graph $torso(G,V^*)$. It follows from the combination of Proposition 2.5.
in \cite{TWRedSTACS} and Proposition \ref{internal} that a set $K$
is the smallest important separator of $G$ if and only if it is the smallest
important separator of $G^*$. Therefore the algorithm first constructs graph $G^*$
and then solves the problem regarding $G^*$.

The algorithm consists of a number of iterations. On the $i$-th iteration the algorithm
either computes a set $S_i$ or returns the answer. The algorithm starts from
setting $S_0=\{y\}$. Assume that the algorithm is in the $i$-th iteration while
it did not return the answer on the $i-1$-th iteration. For $1 \leq j \leq r$,
let $z_j$ be the largest index such that $v_{j,z_j} \notin S_{i-1}$ and let $y_j$ be the
smallest index such that $v_{j,y_j}$ is adjacent to $S_{j-1}$. If for each $j$, $y_j=z_j$,
the algorithm returns the set $\{v_{1,y_1}, \dots v_{r,y_r}\}$. Otherwise, the algorithm
obtains $S_i$ by adding to $S_{i-1}$ the vertices $v_{j,y_j+1}, \dots v_{j,z_j}$ for each
$j$ such that $y_j \neq z_j$.

To analyze the algorithm, observe first that by construction $S_0 \subset S_1 \subset S_2 \dots$
and that for each $S_i$ the subset of each $V(p_j)$ that belongs to $S_i$ forms a suffix
of $p_j$. It follows from the latter statement that each $G^*[S_i]$ is connected.
Furthermore, observe that no $S_i$ intersects with a smallest $X-Y$ separator.
This is certainly true for $S_0$. Assume the truth for $S_{i-1}$. If this is not the case
for $S_i$ then there is a vertex $w$ of a smallest $X-Y$ separator $K'$ that belong to 
the subpath of some $p_j$ whose end vertices are $v_{j,y_{j}+1}$ and $v_{j,z_j}$ as defined
above. It follows that $K'$ does not contain any other vertex of $p_j$. Consequently,
$Y$ can be reached from $X$ in $G^* \setminus K'$ by going along $p_j$ from $x$ to $v_{j,y_j}$
and then jumping to $S_{j-1}$ which is connected and disjoint with $K'$. This contradiction
shows that correctness of the considered observation. It follows that each smallest
$X-Y$ separator is in fact $X-S_i$ separator for all $S_i$ generated during the run of the 
algorithm. Since $S_i$ grows with the increase of $i$, the stopping condition is met after
some $b+1 \leq n$ iterations (i.e. the last constructed set is $S_b$). It is not hard to observe
that the returned set $K$ is a smallest $X-Y$ separator. In fact it is also the desired
important separator. Indeed, by the proven above the component $S_b$ of $Y$ in $G^* \setminus K$ 
is smallest possible in case we consider only smallest $X-Y$ separators. 
Consequently, $NR(G^*,Y,K)=V(G^*) \setminus (K \cup S_b)$
is largest possible. This finishes the correctness proof of the proposed algorithm.

For the runtime, not that $G^*$ can be constructed in $O(n^2)$. 
The $i$-th iteration of the algorithm examines adjacency of $S_{i-1}$
with the rest of the graph. But in fact we can consider only adjacency
of $S_{i-1} \setminus S_{i-2}$ because the only vertices outside $S_{i-1}$
adjacent to $S_{i-2}$ are $v_{1,z_1}, \dots v_{r,z_r}$ known by construction
of $S_{i-1}$, It follows that the adjacency of each pair of vertices is
examined a constant number of times and hence the algorithm takes time $O(n^2)$.
$\blacksquare$

\begin{definition} \label{normalized}
Let $G$ be a graph and $X,Y$ be two disjoint subsets of its vertices.
We say that $G$ is $X-Y$ normalized if $N(X)$ is the only smallest $X-Y$
separator.
\end{definition}

Let $K$ be a $X-Y$ separator.
Denote by $Pr(G,X,Y,K)$ the graph obtained
from $G \setminus (NR(G,Y,K) \setminus X)$ by making $X$ adjacent to
all the vertices of $K$. The graph $PR(G,X,Y,K)$ has the following
easily observable properties.

\begin{proposition} \label{prproperties}
\begin{enumerate}
\item Let $K_1 \geq K $ be an $X-Y$ separator.
Then $K_1$ is a $X-Y$ separator of $Pr(G,K,X,Y)$.
Moreover, if $K_1$ is a smallest $X-Y$ separator of $G$ then
$K_1$ remains a smallest $X-Y$ separator of $Pr(G,K,X,Y)$.
\item Let $K_2 \geq K$ be another $X-Y$ separator.
Then $K_2 \geq K_1$ in $G$ if and only if $K_2 \geq K_1$ in $Pr(G,X,Y,K)$.
In particular, $K_2$ is an important $X-Y$ separator of $G$ if and only if $K_2$
is an important $X-Y$ separator of $Pr(G,X,Y,K)$.
\item If $K$ is an important $X-Y$ separator of $G$ then $Pr(G,X,Y,K)$ is 
$X-Y$ normalized.
\end{enumerate}
\end{proposition}
{\bf Proof.}
For part 1, consider an $X-Y$ path $p$ in $Pr(G,X,Y,K)$.
This path can be transformed into an $X-Y$ path of $G$, possibly
by introducing vertices of $NR(G,Y,K)$. $K_1$ is disjoint with
$NR(G,Y,K)$ by Proposition \ref{internal}. On the other hand,
$K_1$ intersects the transformed path. Consequently, $K_1$ intersects
the initial path $p$. That is, $K_1$ is an $X-Y$ separator of $Pr(G,X,Y,K)$.
Furthermore, since any $X-Y$ separator of $Pr(G,X,Y,K)$ is
clearly an $X-Y$ separator of $G$, any smallest $X-Y$ separator of 
$G$ is also a smallest separator of $Pr(G,X,Y,K)$.

For part 2, apply Proposition \ref{internal} and, arguing as in the 
previous paragraph, observe that $K_2$ separates $K_1 \setminus K_2$
in $G$ if and only if the same happens in $Pr(G,X,Y,K)$. Finally, 
for part 3, observe that if $K$ is not the only smallest separator
of $Pr(G,X,Y,K)$ then $K$ is not important in $Pr(G,X,Y,K)$ in
contradiction to part 2. $\blacksquare$

\section{Important witnesses}

\begin{definition} \label{excess}
Let $G$ be a graph, $X,Y$ be two disjoint subsets of vertices,
$r$ be the smallest size of a $X-Y$ separator and $K$ be an arbitrary $X-Y$ separator.
We call $|K|-r$ the \emph{excess} of $K$ and denote it by $excess_{G,X,Y}(K)$,
the subscripts may be omitted if clear from the context.
\end{definition}

\begin{definition} \label{coverexset}
Let $G$ be a $X-Y$-normalized graph and let $S \subseteq N(X)$. 
We call the excess of a smallest $X-Y$ separator disjoint with $S$
the \emph{cover excess} of $S$ and denote it by $CE_{G,X,Y}(S)$, the subscripts
can be omitted if clear from the context. If $S$ is adjacent to $Y$ then
$CE(S)$ is infinite. A $X-Y$ separator $K$ with $S\cap K=\emptyset$ and $excess(K)=CE(S)$
is called a \emph{witness} of $S$ (w.r.t. $X,Y$ if not clear from the context).
\end{definition}

\begin{lemma} \label{oneimport}
Let $G$ be a $X-Y$-normalized graph and let $S \subseteq N(X)$
and assume that $S$ is not adjacent to $Y$.
There is exactly one important witness $K(S)$ of $S$.
\end{lemma}

{\bf Proof.}
Let $G'$ be the graph obtained from $G$ by splitting each $v \in S$ into $n+1$ copies.
It is not hard to see that $K'$ is a witness of $S$ in $G$ if and only if $K'$
is the smallest separator of $G'$. Furthermore, $K'$, disjoint with $S$, is an important 
$X-Y$ separator of $G$ if and only if
$K'$ is an important separator of $G'$. Combining the above two statements, we conclude
that $K'$ is an important witness of $S$ in $G$ if and only if $K'$ is the smallest
important separator of $G'$. According to Corollary \ref{onesmallest}, there is exactly
one such $K'$. $\blacksquare$

{\bf Remark 1.} If $S=\{v\}$, we write $C(v)$ and $K(v)$ instead of $C(\{v\})$
and $K(\{v\})$, respectively. Also, from now on, 
we will refer to $K(S)$ without special reference
to Lemma \ref{oneimport}.

\begin{lemma} \label{setcover}
Let $G$ be a $X-Y$-normalized graph and let $S \subseteq N(X)$
and assume that $S$ is not adjacent to $Y$. 
Let $K_1$ be an important $X-Y$ separator of $G$ disjoint with $S$ 
and let $K(S)$ be an 
important witness of $S$. Then $K_1 \geq K(S)$.
\end{lemma}

{\bf Proof.}
Let $G'$ be the graph as in the first paragraph of the proof of 
Lemma \ref{oneimport}. Since $K(S)$ is the only smallest important
$X-Y$ separator of $G'$, it follows from Corollary \ref{onesmallest}
that $K' \geq K(S)$ in $G'$. It is not hard to observe that the
same relationship is preserved in $G$. $\blacksquare$

\begin{lemma} \label{manyimport}
Let $G$ be a $X-Y$-normalized graph and let $S \subseteq N(X)$
and assume that $S$ is not adjacent to $Y$. Then there is 
$S' \subseteq S$ such that $|S'| \leq CE(S)$
and $K(S')=K(S)$.
\end{lemma}

{\bf Proof.}
The proof is by induction on $CE(S)$.
Assume first that $CE(S)=1$ and pick an arbitrary vertex $v \in CE(S)$.
We claim that $K(S)=K(v)$. Indeed, according to Lemma \ref{setcover} applied to
$\{v\}$, $K(S) \geq K(v)$. Then, according to Proposition \ref{prproperties},
$K(S)$ is an $X-Y$ separator of $Pr(G,K(v),X,Y)$ and $Pr(G,K(v),X,Y)$ is normalized.
It follows that if $K(S) \neq K(v)$ then $CE(S)=|K(S)|>|K(v)| \geq |N(X)|+1$,
a contradiction. Thus the statement holds in the considered case.

The above reasoning also applies to the case where there is $v \in S$ such
that $CE(v)=CE(S)$. Assume this is not the case. Then we can specify a
maximal $S^* \subseteq S$ such that $CE(S^*)<CE(S)$. By the induction assumption
there is $S'' \subseteq S^*$, $|S''| \leq CE(S^*)$ such that $K(S'')=K(S^*)$.
Pick an arbitrary $v \in S \setminus S^*$. We claim that $K(S)=K(S'' \cup \{v\})$. 
To prove the claim, observe first that $K(S'' \cup \{v\})=K(S^* \cup \{v\})$.
Indeed, according to Lemma \ref{setcover}, $K(S'' \cup \{v\}) \geq K(S'')=K(S^*)$.
It follows that $S^* \cup \{v\} \subseteq N(X) \setminus K(S'' \cup \{v\})$.
Another application of Lemma \ref{setcover} shows that  $K(S'' \cup \{v\}) \geq K(S^* \cup \{v\})$.
On the other hand, $S'' \cup \{v\} \subseteq S^* \cup \{v\}$ and hence, yet another application
of Lemma \ref{setcover} implies $K(S^* \cup \{v\}) \geq K(S'' \cup \{v\})$, yielding the desired
equality. Now, observe that $K(S)=K(S^* \cup \{v\})$. Indeed, by Lemma \ref{setcover},
$K(S) \geq K(S^* \cup \{v\})$. On the other hand, due to the minimality of $S^*$,
$K(S) \ngtr K(S^* \cup \{v\})$. The claim now follows. 
$\blacksquare$

\section{Upper bound on the number of important separators and the \textsc{mwc} problem}
Let $G$ be an $X-Y$ normalized graph
$(S_1, \dots S_r)$ be a sequence of disjoint non-empty subsets of vertices of $G$
and $K$ is an $X-Y$ separator. We say  that $K$ is a \emph{compound witness} of 
the \emph{attribute} $(S_1, \dots, S_r)$ (w.r.t. $X$ and $Y$ in $G$ if clarification is needed)
as follows.  Assume first that $r=1$. Then
$S_1 \subseteq N(X)$ and $K=K(S_1)$. 
Otherwise, $S_2 \cup \dots \cup S_r$ is disjoint with $N(X)$ and $K$ is a compound witness
of $(S_2, \dots, S_r)$ w.r.t. $X,Y$ in $Pr(G,X,Y,K(S_1))$. 
We call $|S_1|+ \dots + |S_r|$ the \emph{rank} of $K$. The following
corollary immediately follows from inductive application of Lemma \ref{oneimport}.

\begin{corollary} \label{manydimimport}
Each sequence $(S_1, \dots, S_r)$ is the attribute of \emph{at most one} compound 
witness. (Some sequences may correspond to no compound witness, for example, due to being
non well-formed attributes.)
\end{corollary}

\begin{theorem} \label{ComputeWitness}
Let $G$ be a $X-Y$ normalized graph and $(S_1, \dots, S_r)$ be a sequence
of disjoint non-empty sets of vertices. Then the existence of a compound witness
with attribute $(S_1, \dots, S_r)$ can be tested in $O(n^3)$.
\end{theorem}

{\bf Proof.}
Consider the following algorithm. First, compute the unique smallest important separator $K_0$ of $G_0=G$. 
Then obtain graph $G_1$ by introducing extra copies of vertices of $S_1$ in $Pr(G_0,X,Y,K)$
and compute the smallest important separator $K_1$. Then obtain graph $G_2$
from $Pr(G_1,X,Y,K_1)$ by introducing extra copies of vertices of $S_2$ and so on
until $K_r$ is eventually returned. The algorithm can also return 'NO' if some
$X$ or some intermediate $K_i$ is adjacent to $Y$ or if some $S_i$ is not a subset of $K_{i-1}$.
The correctness of this algorithm follows from definition of the attribute. 

The runtime $O(rn^3)$ immediately follows from Lemma \ref{ComputeImport}.
However, using an amortisation
argument we can show that in fact $O(n^3)$ is enough. Denote $|K_i|$ by $z_i$
and assume w.l.o.g. that $K_r$ is successfully computed (otherwise we can
consider computation until some $K_{r'}$ for $r'<r$. By Proposition \ref{prproperties},
$z_0< \dots <z_r$. Now, consider graph $G_1$. 
It is not hard to see that the $z_0$ internally vertex disjoint
$X-Y$ paths of $G_0$ (found during the run of network flow algorithm) 
are naturally transformed into $z_0$ internally vertex disjoint $X-Y$ paths of
$G_1$.These paths provide initial flow of size $z_0$ and hence only
$(z_1-z_0)+1$ additional iterations of the Ford-Fulkerson algorithm will be
needed for the next iteration of the algorithm of Lemma \ref{ComputeImport} to produce  
the largest set of internally vertex disjoint $X-Y$ path of $G_1$. 
Applying this argument inductively, it is not hard to observe that
the resulting algorithm takes $O(n)$ iterations of Ford-Fulkerson algorithm.
Each of these iterations takes $O(n^2)$. In addition there are at most $n$
iterations of computing the smallest important separator, each requiring $O(n^2)$
time according to Lemma \ref{ComputeImport}. Finally, the algorithm also creates
a $Pr$-graph at most $n$ times, $O(n^2)$ per creation is clearly enough. 
Consequently, the overall runtime is $O(n^3)$. $\blacksquare$

\begin{theorem} \label{ImportCompound}
Let $G$ be a $X-Y$-normalized graph and let $K \neq N(X)$ an important $X-Y$
separator. Then $K$ is a compound witness of rank at most $excess(K)$.
\end{theorem}

{\bf Proof.}
By induction on $excess(K)$. Assume first that $excess(K)=1$ and let $v \in N(X) \setminus K$.
Then $K=K(v)$, as shown in the first paragraph of proof of Lemma \ref{manyimport}.
In other words, in the considered case, $K$ is a compound witness with attribute $(\{v\})$.

Assume now that $excess(K)>1$. Denote $N(X) \setminus K$ by $S$. According to Lemma \ref{setcover},
$K \geq K(S)$. Furthermore, according to Lemma \ref{manyimport}, there is $S_1 \subseteq S$ with
$|S_1| \leq CE(S)$ such that $K(S)=K(S_1)$. If $K=K(S)$ then $(S_1)$ is the desired attribute.
Otherwise, denote $P(G,X,Y,K(S))$ by $G_1$. According to Proposition \ref{prproperties},
$G_1$ is normalized and $K$ is an important $X-Y$ separator of $G_1$. Furthermore, 
$excess_{G_1,X,Y}K =excess_{G,X,Y}K-CE(S_1)<excess_{G,X,Y}(K)$. By the induction assumption, $K$ is a compound witness
w.r.t. $X,Y$ in $G_1$ of rank at most $excess_{G_1,X,Y}K$. Let $(S_2, \dots, S_r)$ be the 
corresponding attribute. We claim that $K$ is the compound witness of $(S_1, \dots, S_r)$
w.r.t. $X,Y$ in $G$. Indeed, $|S_1|+ \sum_{i=2}^{r} |S_i| \leq CE(S_1)+ excess_{G_1,X,Y}K=excess_{G,X,Y}K$,
the inequality is obtained by definition of $S_1$ and the induction assumption,
the equality is obtained by definition of $G_1$.

It remains to show that $S_2, \dots, S_r$ are disjoint with $N(X)$. First of all, 
note that $K$ is disjoint with $S_1$. Furthermore, inductively applying the definition of 
a compound witness, it is not hard to see that $K$ is disjoint with $S_2, \dots, S_r$.
Since each of $S_2, \dots, S_r$ are subsets of vertices of $Pr(G,X,Y,K(S))$, they
are all disjoint with $S$. It follows that if some $S_i$ is not disjoint with
$N(X)$, it is in fact not disjoint with $N(X) \setminus S$. Let $v \in (N(X) \setminus S) \cap S_i$.
It follows that $v \notin K$ in contradiction to $N(X) \setminus K=S$. $\blacksquare$


\begin{theorem} \label{MainBound}
Let $G$ be a graph and let $X$ and $Y$ bet two non-intersecting subsets of $V(G)$.
Let $k>0$ be an integer. Then there are at most $\sum_{i=0}^k{n \choose i}$ important $X-Y$ separators
of excess at most $k$. Moreover, they can be generated by considering all subsets
of at most $k$ vertices of $G$ with an $O(n^3)$ time spent per subset.
\end{theorem}

{\bf Proof.}
First of all we show that we can assume that $G$ is an $X-Y$ normalized graph.
Indeed, assume that $G$ is not such graph and let $K^*$ be the only smallest
important separator existing according to Corollary \ref{onesmallest}. 
Let $K'$ be an arbitrary important separator. According to 
Corollary \ref{onesmallest}, $K' \geq K^*$. 
It follows from Proposition \ref{prproperties}
that the set of important $X-Y$ separators of $G$ is the same as the set of
important $X-Y$ separators of $Pr(G,X,Y,K^*)$ and that $Pr(G,X,Y,K^*)$ is normalized.
This shows the validity of assumption that $G$ is an $X-Y$ normalized graph.
The ${n \choose 0}$ in the claimed bound stands for the unique smallest important $X-Y$ separator, $N(X)$ in
our case. We are now going to show that the number of the rest of important $X-Y$ separators
is at most  $\sum_{i=1}^k{n \choose i}$.

Let us say that a set $S$ \emph{corresponds} to an attribute $(S_1, \dots, S_r)$ (and vice versa
the attribute corresponds to the set)
if  $\bigcup_{i=1}^r S_i=S$. We show that each subset $S$ of $V(G)$ 
corresponds to at most one well-formed attribute 
$(S_1, \dots S_r)$  of a compound witness. The proof is by induction.
The empty set does not correspond to any well-formed attribute.
Assume that $|S|=1$. If $S$ is disjoint with $N(X)$ then again $S$ does not
correspond to any well-formed attribute. Otherwise, $S \subseteq N(X)$
and the only attribute $S$ can correspond to is $(S)$.
Assume now that $|S|>1$. If $S$ is disjoint with $N(X)$ then once again
$S$ does not correspond to any well-formed attribute. Otherwise,
let $(S_1, \dots, S_r)$ be an attribute corresponding to $S$. Observe that
$S_1=S \cap N(X)$. Furthermore, by the induction assumption,
 $(S_2, \dots, S_r)$ is the unique attribute corresponding to $S \setminus S_1$.
Taking into account the uniqueness of $S_1$, the uniqueness of $(S_1, \dots, S_r)$
follows. 

The correspondence established above tells us that there are at 
most $\sum_{i=1}^k{n \choose i}$ well-formed
attributes of rank at most $k$. Since according to Corollary \ref{manydimimport}, each
$(S_1, \dots, S_r)$ is the attribute of at most one compound witness w.r.t. $X$ and $Y$,
the number of compound witnesses of rank at most $k$ is also bounded by $\sum_{i=1}^k{n \choose i}$. 
Theorem \ref{ImportCompound} implies the same bound on the number of important 
$X-Y$ separators different from $N(X)$ and having excess at most $k$. 
Finally, the runtime upper bound follows from Theorem \ref{ComputeWitness}. 
$\blacksquare$

With Theorem \ref{MainBound} in mind we are ready to compute the runtime of solving \textsc{mwc}
problem. Let $(G,T)$ be an instance of the multiway cut problem where $G$ is a graph and $T$ is the set
of terminals to be separated. Let $t \in T$. We call a $t-T \setminus t$ separator of $G$
an \emph{isolating cut} of $t$ (w.r.t. $(G,T)$ if the context is not clear).
The following lemma has is a reformulation of Lemma 3.6. of \cite{MarxTCS}.

\begin{lemma}\label{IsoImport}
For any $t \in T$ there is an optimal solution of $(G,T)$
containing an important isolating cut of $t$.
\end{lemma}

\begin{theorem}
Let $(G,T)$ be an instance of the multiway cut problem.
For $t \in T$, let $m(t)$ be the size of the smallest isolating cut
of $t$. Let $m=max_{t \in T} m(t)$ and $s$ be an integer. 
Then there is $O(sn^{s+3}+|T|n^3)$ algorithm that checks whether $(G,T)$ 
has a solution of size at most $(m+s)$.
\end{theorem}

{\bf Proof.}
For each terminal of $T$ compute the respective smallest important
isolating cut. According to Lemma \ref{ComputeImport}, this can be
done in $O(n^3)$ per terminal, so the overall time spent in $O(|T|n^3)$.
Let $t$ be the terminal whose respective smallest important isolating cut 
is of size $m$. If $k=0$ then, according to Lemma \ref{IsoImport} and
Corollary \ref{onesmallest}, either this isolating cut is the solution
or there is no solution.

If $k>0$, the algorithm generates all possible important isolating cuts $K$ of $t$ of excess 
at most $k$. For each such $K$, it solves the instance $(G \setminus K,T \setminus \{t\},m+k-|K|)$ and 
returns 'YES' if and only if at least one such residual instance has a solution.
The correctness of this approach follows from Lemma \ref{IsoImport}.
Furthermore, since $|K| \geq m$, $m+k-|K| \leq k$.

According to \cite{ChenMwayCut},
each residual instance can be solved in time $O(n^3(k-i)4^{k-i})$, where $i$ is the excess of $K$. 
According to Theorem \ref{MainBound}, for each $i \leq k$ there are at most $\sum_{j=0}^i{n \choose i}$
important isolating cuts of $t$ of excess $i$. Moreover, they can be enumerated by spending
$O(n^3)$ for each of them. The proposed approach requires to spend additional
time $O(n^3(k-i)4^{k-i})$ per isolating cut of $t$. The overall time spent per an isolating cut of 
$t$ is thus  $O(n^3(k-i)4^{k-i}+n^3) \subseteq O(n^3k4^{k-i})$.
Taking into account that $\sum_{j=0}^i{n \choose j} \leq c{k \choose i}n^i$ for some constant $c$
the resulting runtime is $O(n^3k \sum_i {k \choose i}n^i4^{k-i})=O(n^3k(n+4)^k)$. The desired runtime
can be obtained by taking into account that $(n+4)^k$ and $n^k$ are asymptotically the same.
$\blacksquare$


\begin{thebibliography}{10}

\bibitem{MedImaging}
Yuri Boykov and Marie-Pierre Jolly.
\newblock Interactive organ segmentation using graph cuts.
\newblock In {\em MICCAI}, pages 276--286, 2000.

\bibitem{Zabih1}
Yuri Boykov, Olga Veksler, and Ramin Zabih.
\newblock Fast approximate energy minimization via graph cuts.
\newblock {\em IEEE Trans. Pattern Anal. Mach. Intell.}, 23(11):1222--1239,
  2001.

\bibitem{Rabaniapprox}
Gruia Calinescu, Howard~J. Karloff, and Yuval Rabani.
\newblock An improved approximation algorithm for multiway cut.
\newblock {\em Journal of Computer and System Sciences}, 60(3):564--574, 2000.

\bibitem{ChenMwayCut}
Jianer Chen, Yang Liu, and Songjian Lu.
\newblock An improved parameterized algorithm for the minimum node multiway cut
  problem.
\newblock In {\em WADS}, pages 495--506, 2007.

\bibitem{ChenAlgorithmica}
Jianer Chen, Yang Liu, and Songjian Lu.
\newblock An improved parameterized algorithm for the minimum node multiway cut
  problem.
\newblock {\em Algorithmica}, 55(1):1--13, 2009.

\bibitem{DFVSalgo}
Jianer Chen, Yang Liu, Songjian Lu, Barry O'Sullivan, and Igor Razgon.
\newblock A fixed-parameter algorithm for the directed feedback vertex set
  problem.
\newblock {\em Journal of the ACM}, 55(5), 2008.

\bibitem{Cormen}
Thomas Cormen, Charles~E. Leiserson, Ronald~L. Rivest, and Clifford Stein.
\newblock {\em Introduction to Algorithms}.
\newblock The MIT Press, 3nd edition, 2009.

\bibitem{Dahlapprox}
Elias Dahlhaus, David~S. Johnson, Christos~H. Papadimitriou, Paul~D. Seymour,
  and Mihalis Yannakakis.
\newblock The complexity of multiterminal cuts.
\newblock {\em SIAM J. Comput.}, 23(4):864--894, 1994.

\bibitem{MwayCutTrees}
P{\'e}ter~L. Erd{\"o}s and L{\'a}szl{\'o}~A. Sz{\'e}kely.
\newblock On weighted multiway cuts in trees.
\newblock {\em Math. Program.}, 65:93--105, 1994.

\bibitem{Grohebook}
J{\"o}rg Flum and Martin Grohe.
\newblock {\em Parameterized Complexity Theory (Texts in Theoretical Computer
  Science. An EATCS Series)}.
\newblock Springer-Verlag, 2006.

\bibitem{Gargapprox}
Naveen Garg, Vijay~V. Vazirani, and Mihalis Yannakakis.
\newblock Multiway cuts in node weighted graphs.
\newblock {\em Journal of Algorithms}, 50(1):49--61, 2004.

\bibitem{GutinESA}
Gregory Gutin, Leo van Iersel, Matthias Mnich, and Anders Yeo.
\newblock All ternary permutation constraint satisfaction problems
  parameterized above average have kernels with quadratic numbers of variables.
\newblock In {\em ESA (1)}, pages 326--337, 2010.

\bibitem{Hartvigsen}
David Hartvigsen.
\newblock The planar multiterminal cut problem.
\newblock {\em Discrete Applied Mathematics}, 85(3):203--222, 1998.

\bibitem{LokshtanovClustering}
Daniel Lokshtanov and D{\'a}niel Marx.
\newblock Clustering with partial information, 2010.

\bibitem{RamanAbove}
Meena Mahajan and Venkatesh Raman.
\newblock Parameterizing above guaranteed values: Maxsat and maxcut.
\newblock {\em J. Algorithms}, 31(2):335--354, 1999.

\bibitem{MarxTCS}
D{\'a}niel Marx.
\newblock Parameterized graph separation problems.
\newblock {\em Theor. Comput. Sci.}, 351(3):394--406, 2006.

\bibitem{TWRedSTACS}
D{\'a}niel Marx, Barry O'Sullivan, and Igor Razgon.
\newblock Treewidth reduction for constrained separation and bipartization
  problems.
\newblock In {\em STACS 2010}, pages 561--572, 2010.

\bibitem{Multicut}
D{\'a}niel Marx and Igor Razgon.
\newblock Fixed-parameter tractability of multicut parameterized by the size of
  the cutset.
\newblock {\em CoRR}, abs/1010.3633, 2010.

\bibitem{ROJCSS}
Igor Razgon and Barry O'Sullivan.
\newblock Almost 2-sat is fixed-parameter tractable.
\newblock {\em Journal of Computer and System Sciences}, 75:435--450, 2009.

\bibitem{StoneApplication}
Harold Stone.
\newblock Multiprocessor scheduling with the aid of netowrk flow algorithms.
\newblock {\em {I}{E}{E}{E} Transactions on Software Engineering}, 1:85--93,
  1977.

\end{thebibliography}
\end{document}